\documentclass[12pt,oneside]{article}
\usepackage{amsfonts,amssymb,graphicx}
\input{epsf.tex}
\def\be{\begin{equation}}
\def\ee{\end{equation}}
\def\bea{\begin{eqnarray}}
\def\eea{\end{eqnarray}}
\def\<{\langle}
\def\>{\rangle}
\def\~{\tilde}

\def\blackbox{\rightline{\vrule height 1.7ex width 1.2ex depth -.5ex}}

\newcommand{\Z}{\Bbb Z}

\newtheorem{theorem}{Theorem}
\newtheorem{lemma}{Lemma}
\newtheorem{definition}{Definition}
\newtheorem{corollary}{Corollary}

\begin{document}
\begin{center}
{\bf\sc\Large
Path Integral Representations for the\\ 
Spin-Pinned Quantum XXZ Chain }\\
\vspace{1cm}
{\linespread{1}
{Oscar Bolina\footnote{Department of Mathematics, University
of California, Davis, CA 95616, USA \\
Permanent address: Departamento de Fisica-Matematica,
Universidade de Sao Paulo, Sao Paulo 05315-970 Brazil}, 
Pierluigi Contucci\footnote{Dipartimento di Matematica,
Universit\`a di Bologna, 40127 Bologna, Italy}, Bruno 
Nachtergaele \footnote{Department of Mathematics, University
of California, Davis, CA 95616, USA}}\\
\vspace{.5cm}
oscar@math.ucdavis.edu, contucci@dm.unibo.it, bxn@math.ucdavis.edu\\ 
}%reset linespread
\vskip 1truecm
{\small June 3th 2003, revised September 18th 2003}
\end{center}
\vskip 1truecm
\begin{abstract}\noindent
Two discrete path integral formulations for the ground state of a spin- 
pinned quantum anisotropic XXZ Heisenberg chain are introduced.  Their 
properties are discussed and two recursion relations are proved.
\end{abstract}
\newpage

\section{Introduction}

We introduce in this work a path integral representation for the ground state
of the anisotropic Heisenberg XXZ model with a pinned-spin as a suitable random
walk on two dimensional lattices. Our representation generalises what we had
previously introduced for the standard XXZ chain. The reason to introduce
spin-pinned models is to deal with localised impurities in magnetic materials
\cite{CNS,NS,Starr}. The path integral representation is of great help in
establishing the properties of the model under investigation: the square norm
of the quantum state vector admits the interpretation of the path integral
partition function, and the probabilistic features, in particular related to
Markov type properties, play a decisive role in evaluating correlation functions
and other physical quantities. We prove here two recursion relations that
express the properties of systems of a given size in terms of those of smaller
size. Recursion relations of this kind have been used successfully to derive
bounds on correlation functions in \cite{BCN}.

\section{Path Integral Models in $\Z^2$}

Let us consider the two dimensional lattice ${\bf Z}^2$. A ``zig-zag'' path $p$ 
is a connected path of unit steps 
in ${\bf Z}^2$ monotonically increasing in both coordinates. For each 
$t=1,2,3...$ the 
path will be encoded
in a sequence of  $\alpha(t)\in \{0,1\}$ conventionally associating 
$\alpha=1$ to a 
horizontal step and $\alpha=0$ to a vertical one. We denote by $|p|$ 
the length of the path
i.e. the sum of the steps (see Figure 1). 
\begin{figure}
\centerline{\epsfbox{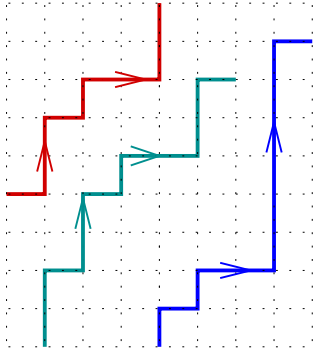}}
\caption{Examples of zig-zag paths on $\bf{Z}^{2}$.}
\end{figure}
\noindent
A path integral model on $\bf{Z}^2$ is a law that associates
positive weights $w(p)$ to a given set of paths.
\newline
Let ${\cal P}_{I,F}$  denote the set of all paths
from an initial point $I=(n_0,m_0)$
to a final one $F(n_f,m_f)$ for $n_0\le n_f$ and $m_0\le m_f$. 
The collection of all such paths visits all the points of the {\it
rectangle} $[I,F]$. The {\it canonical} partition function is defined as
\be\label{GN}
Z(I,F) \; = \; \sum_{p \in {\cal P}_{I,F}} w(p) \; ,
\ee
which induce the probability measure on ${\cal P}_{I,F}$ given by
\be\label{pr}
{\rm Prob}(p) \; = \; \frac{w(p)}{Z(I,F)} \; .
\ee
In path integral models, correlation functions measure the probability
that a path goes through particular points $Q_1=(i_1,j_1), Q_2=(i_{2},j_{2}), 
\cdots, Q_r=(i_{r},j_{r})$, with  
\be
n_o \leq i_1 \leq i_2 \leq ... \leq n_f \; ,
\ee
and
\be
m_o \leq j_1 \leq y_2 \leq ... \leq m_f \; .
\ee
The one-point correlation function is defined
as the probability of crossing a single point $Q$
\be\label{prob1}
P_{I,F}(Q)=\frac{Z(I,F \mid Q)}{Z(I,F)} \; ,
\ee
where
\be
Z(I,F \mid Q) \; = \; \sum_{p\in {\cal P}_{I,F}(Q)} w(p)
\label{condpar}
\ee
and ${\cal P}_{(I,F)}(Q)$ is the set of paths from the $I$ to
$F$ that pass through the point $Q$. More generally, we can 
define
\be\label{probr}
P_{I,F}(Q_1;\cdots;Q_r)=
\frac{Z(I,F \mid Q_1; \cdots; Q_r)}{Z(I,F)},
\ee
where
\be
Z(I,F \mid Q_1; \cdots; Q_r)
\; = \; \sum_{p\in {\cal P}_{I,F}(Q_1; \cdots; Q_r)} w(p)
\label{condparr}
\ee
and ${\cal P}_{I,F}(Q_1; \cdots; Q_r)$ denotes the set of 
paths that pass through the particular points $Q_1, Q_2, 
\cdots, Q_r$. 
\newline
In this framework, we consider models for which the weight $w(p)$
is a local function of the bonds that the path is passing through. 
Denoting by ${\bf B}^{2}$ the set of bonds in ${\bf Z}^2$, 
we associate a positive number $w(b)$ to each element $b$ of 
${\bf B}^{2}$ and define
\be\label{local}
w(p)=\prod_{b \in p} w(b).
\ee
More generally for a given finite set of paths ${\cal P}$ (the {\it 
ensemble}) we define
\be\label{GGN}
{\cal Z} \; = \; \sum_{p \in {\cal P}} w(p) \; , 
\ee
for a set of paths ${\cal P}(Q)$ through a point $Q$, 
\be\label{GN1}
{\cal Z}(Q) \; = \; \sum_{p \in {\cal P}(Q)} w(p) \; 
\ee
and for a set of paths ${\cal P}^{(+)}(Q)$ (resp. ${\cal P}^{(-)}(Q)$) 
{\it ending} (resp. {\it beginning}) in $Q$, 
\be\label{GGN1}
{\cal Z}^{(\pm)}(Q) \; = \; \sum_{p \in {\cal P}^{(\pm)}(Q)} w(p) \; .
\ee
\\
In order to prove the following basic {\it Markov} property we show the
following lemma.
{\lemma [Markov property]
\be\label{N1kj} 
Z(I,F \mid Q_1; \cdots; Q_r)\; = \; Z(I,Q_1)Z(Q_1,Q_2)\cdots Z(Q_r,F)\; 
\ee
and analogously
\be
{\cal Z}(Q) \; = \; {\cal Z}^{(+)}(Q){\cal Z}^{(-)}(Q)
\ee 
}
\begin{itemize}
\item[] {\bf Proof.} These identities follow from the fact that the paths
are increasing in both coordinates and from (\ref{local}).
\newline 
\blackbox
\end{itemize}
{\corollary \label{l1}
Let the set ${\cal S}_{I}(l)$ (sphere of center $I$ and radius $l$) be 
the points reachable by the paths $p$ starting at
$I$  of lenght $l$. For any {\it sphere}
${\cal S}_{I}(l)$ such that $l\le n_f+m_f-n_0-m_0$  we have 
\be
Z(I,F) \; = \; \sum_{Q\in {\cal S}_{I}(l)} Z(I,Q)Z(Q,F)
\label{genrec}
\ee
}
\noindent
\begin{itemize}
\item[] {\bf Proof.} We write (\ref{GN}) with the set of paths  
${\cal P}_{(I,F)}=\cup_{Q\in {\cal S}_I(l)} {\cal P}_{I,F}(Q)$
as
\be\label{GNN}
Z(I,F) \; = \; \sum_{ \bigcup_{Q\in {\cal S}_I(l)} {\cal P}_{I,F}(Q)} 
\sum_{p\in {\cal P}_{I,F}(Q)} w(p). 
\ee
and by lemma (\ref{l1}) we have the corollary.
\newline 
\blackbox
\end{itemize}
In a completely analogous way the following can be proved:
{\corollary \label{l2}
Let $b_h$ and $b_v$ the two bonds leading to (resp. departing from) $Q$ 
with $b_h=(Q_h,Q)$ and
$b_v=(Q_v,Q)$. Then
\be
{\cal Z}^{(\pm)}(Q)  \; = \; w(b_h){\cal Z}^{(\pm)}(Q_h)+w(b_v){\cal 
Z}^{(\pm)}(Q_v)
\label{genrecq}
\ee
}
\section{The XXZ Spin-Pinned Chain}
\noindent
In one dimension, for $0 < q < 1$, we consider the Hamiltonian
\be\label{ham}
H^{}_{[-L,K]}=\sum_{x=-L}^{-1} h^{(q^{-1})}_{x} + \sum_{x=0}^{K-1} h^{(q)}_{x},
\ee
where 
\be
h^{(q)}_{x}=-\frac{2}{q+q^{-1}} (S^{(1)}_{x} S^{(1)}_{x+1}+S^{(2)}_{x}
S^{(2)}_{x+1})-(S^{(3)}_{x} S^{(3)}_{x+1}-1/4)
-\frac{q^{-1}-q}{2(q^{-1}+q)}(S^{(3)}_{x} - S^{(3)}_{x+1})
\ee
is the orthogonal projection on the vector
\be
\xi_{q}=\frac{1}{\sqrt{1+q^{2}}} (q \mid \uparrow \downarrow \rangle-
\mid \downarrow \uparrow \rangle).
\ee
and $S^{(i)}_{x}$ ($i=1,2,3$) are the usual Pauli spin matrices at
the site {\it x}. From the definition of $\xi_{q}$ if follows that
\be
h^q \mid \downarrow \downarrow \rangle = 0, \;\;\;\;\;\;\;\;\;\;
h^q \mid \downarrow \uparrow \rangle = {1 \over {q + q^{-1}}}
\left( q \mid \downarrow \uparrow \rangle - \mid \uparrow \downarrow
\rangle  \right), \;\;\;\;\;\;
\label{dd}
\ee
\be
h^q \mid \uparrow \uparrow \rangle = 0, \;\;\;\;\;\;\;\;\;\;\;
h^q \mid \uparrow \downarrow \rangle = - {1 \over {q + q^{-1}}}
\left( \mid \downarrow \uparrow \rangle - q^{-1} \mid \uparrow 
\downarrow  \rangle \right).
\label{uu}
\ee
\begin{figure}\label{path1}
\centerline{\epsfbox{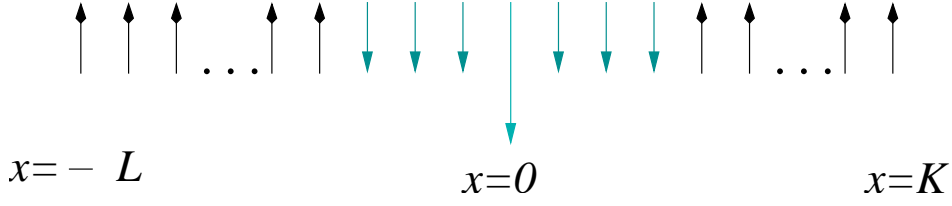}}
\caption{The spin-pinned chain.}
\end{figure}
\noindent
The Hamiltonian (\ref{ham}) represents a spin-$1/2$ XXZ ferromagnetic chain (see
\cite{ASW,GW}) of lenght $K+L+1$ with a kink-antikink structure: two $(+)$ boundary
fields and a spin pinned at the origin through the action of a local $(-)$ field
whose strenght is twice the one on the boundaries (see Figure 2). The configuration of spins in
the one-dimensional chain is identified with the set of numbers ${\alpha_x}$ for $x=\{-L,
..., K\}$ where 
$\alpha$ takes values in $\{0,1\}$. 
We choose $\alpha=0$ 
to correspond to an up spin, or, in the particle language, to an
unoccupied site. Conversely, $\alpha=1$ corresponds to a down spin 
or an occupied site. We will be interested in the ground state of the 
model in the sector with {\it N} down spins. 
We have the following result.
{\theorem A ground state of the model is given by
\be\label{gs}
\psi_N(-L,K)=\sum_{\{\alpha_{x}\}\in {\cal A}_{N,M} }
\phi(\{\alpha_{x} \}) \mid \{ \alpha_{x} \} \rangle
\ee
where the ${\cal A}_{N,M}$ the set of configurations $\{ \alpha_{x} \}$
such that $\sum_{x} \alpha_{x}=N$ with the condition $N+M=L+K+1$, and the functions $\phi(\alpha)$
satisfy the set of equations
\begin{eqnarray}\label{soe}
\phi(..., ~\sigma_{x}=\uparrow, ~\sigma_{x+1}=\downarrow,~ ...)&=&q^{-1}
\phi(..., ~\sigma_{x}=\downarrow, ~\sigma_{x+1}=\uparrow,~ ...) \;\;\;{\rm
for} \;\;\; x < 0, \nonumber \\
\phi(..., ~\sigma_{x}=\uparrow, ~\sigma_{x+1}=\downarrow,~ ...)&=&q~
\phi(..., ~\sigma_{x}=\downarrow, ~\sigma_{x+1}=\uparrow,~ ...)
\,\,\;\;\;\; {\rm for} \;\;\; x \geq 0. 
\end{eqnarray}
}
\begin{itemize}
\item[] {\bf Proof.} Follows by direct substitution of (\ref{gs}) in 
(\ref{ham}) using (\ref{dd}) and (\ref{uu}).
\newline
\blackbox
\end{itemize}
{\theorem The function 
\be\label{sol}
\phi(\alpha)=\prod_{x=-L}^{K} q^{\mid x \mid \alpha_{x}}
\ee
is the solution of (\ref{soe}).
}
\begin{itemize}
\item[] {\bf Proof.} Since the set of equations (\ref{soe}) admits a 
unique solution in each sector of fixed down spins, we are left with 
proving that (\ref{sol}) satisfies
\be\label{sol1}
\frac{\phi(\cdots,~\alpha_{x}=1, ~\alpha_{x+1}=0, ~\cdots)}{\phi(\cdots,~
\alpha_{x}=0,~\alpha_{x+1}=1, ~\cdots)} = 
\left \{
\begin{array}{lll}
q & {\rm when\ } x <0, \\
q^{-1} & {\rm when\ } x \geq 0 \\
\end{array}
\right.
\ee
Since (\ref{sol1}) equals 
\be
\frac{q^{\mid x \mid}}{q^{\mid x+1 \mid}},
\ee
the proof is complete.
\newline 
\blackbox
\end{itemize}
The norm of the ground state vector (\ref{gs}) with {\it n} spins down is
\be\label{norm}
\| \psi_N(-L,K) \|^2=\sum_{\{\alpha_{x}\}\in{\cal A}_{N,M}}~
\prod_{x=-L}^{K}\, q^{2|x| \alpha_{x}}.
\ee

\section{Two Path Integral Representations.}

The two path integral representations that we introduce here are based on
the one introduced in \cite{BCN} for the anisotropic XXZ quantum chain 
(with no pinning field), which we first recall here for completeness. 

{\theorem [Path integral representation for interface ground 
state \cite{BCN}]\label{main}
\be\label{N111}
Z_q(n,m)\; = \;\sum_{\{\alpha_{x}\}\in {\cal A}_{n,m}}~
\prod_{x=1}^{K}\, q^{2x \alpha_{x}}\; =\; \sum_{p \in {\cal P}_{(n,m)}} 
w(p) \; 
\ee
is the partition function for the classical path integral 
model associated with the quantum {\it XXZ} model with $n$ down spins and 
$m$ up spins ($n+m=K$) for the following choice of weights
\be\label{bond1}
w(b)=\left \{
\begin{array}{ll}
q^{2(i_b+j_b)} \;\;\; 
{\rm for~ a~ horizontal~ bond~ whose~ right~ end~ is~ at}~ (i_b,j_b) \\
1 \;\;\;\;\;\;\;\;\;\;\;\;\;\; {\rm any~ vertical~ bond} \; .
\end{array}
\right.
\ee
The partition function (\ref{N111}) has the following explicit expression:
\be\label{ex}
Z_q(n,m)\; =\; q^{n(n+1)}~ \frac{\prod_{i=1}^{n+m} (1-q^{2i})}{
\prod_{i=1}^{n} (1-q^{2i})~ \prod_{i=1}^{m} (1-q^{2i})}.
\ee
Moreover for every $I=(n',m')$, $F=(n,m)$ and $P=(x,y)$ with $x\le n'\le n$ and $y\le m'\le
m$
\be\label{TF}
Z_q(I,F) \, = \, q^{2(x+y)(n-n')} Z_q(I',F'),
\ee
where $I'=(n'-x,m'-y)$, and $F'=(n-x,m-y)$.
}
\vskip .1 cm
\noindent
Next, we state the definitions of two path integral representations,
i.e., two path measures. Although the measures are different, it will
turn out that both measures generate the same family of partition
functions.

{\definition [Path Integral representation 1]
To each configuration of $\alpha\in{\cal A}_{N,M}$ representing a spin 
configuration for the chain in $[-L,K]$ we associate a path $p(\alpha)$ 
starting from the origin of the lattice and ending at $N,M$ described by 
the sequence of $\alpha(t)$, $1\le t \le L+K+1$ \\
\be\label{bond2}
\alpha(t)=\left \{
\begin{array}{cccc}
\alpha_t & ~{\rm for}~ & 1 \le t \leq K \\
\alpha_{t-L-K-1} & ~{\rm for}~ &  K\le t \le K+L+1 \; ,
\end{array}
\right.
\ee
\\
and consider the weights system defined by
\be\label{bond3}
w(b)=\left \{
\begin{array}{cccc}
q^{2(i_b+j_b)} & ~{\rm for}~ & i_{b}+j_{b} \leq K \\
q^{2(K+L+1)-2(i_b+j_b)} & ~{\rm for}~ &  K\le i_{b}+j_{b} \le K+L+1 \\
1 & ~{\rm for}~ &{\rm any~vertical~ bond}. \;\;\;\;\;\;\;\;\;\;\; 
\end{array}
\right.
\ee
We will denote by $Z(N,M)$ the partition function corresponding to the 
given weights and the set of paths ${\cal P}_{(0,0),(N,M)}$. See Figure 3.} 
\begin{figure}
\centerline{\epsfbox{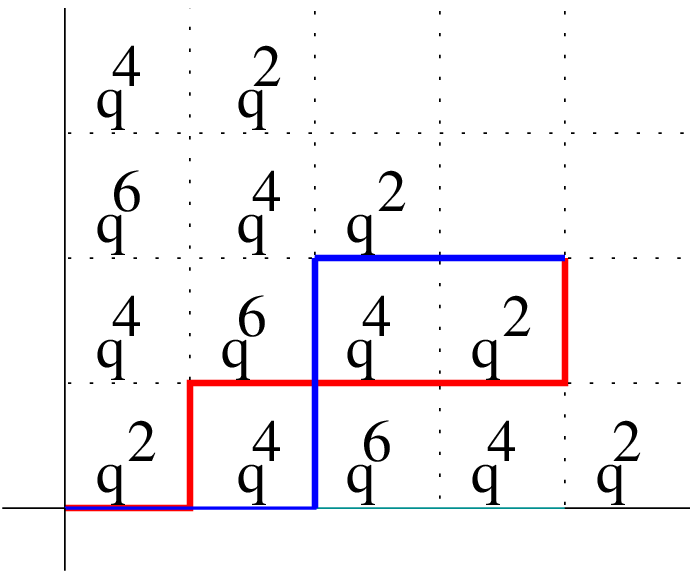}}
\caption{The first path representation showing two paths. 
The weights (\ref{bond3}) are indicated on the bonds. Note that 
the weights are constant along lattice spheres of center $(0,0)$ 
and radius $l$ for $1 \leq l \leq N+M$.
}
\end{figure}

{\definition [Path Integral representation 2]
In this case the spin configuration will correspond to a path 
\be\label{bond300}
\alpha(t) \; = \; \alpha_{t-L-1} 
\ee
and consider the weights system defined by
\be\label{bond30}
w(b)=\left \{
\begin{array}{cccc}
q^{2|i_b+j_b|}  ~&{\rm for}& {\rm ~horizontal~ bonds}  \\
1 ~&{\rm for}& {\rm ~vertical~ bonds}. 
\end{array}
\right.
\ee
The set of paths $\widetilde{\cal P}_{0}(N,M)$ is the set containing
all paths departing from the third quadrant sphere of radius $L+1$, 
and arriving at the first quadrant sphere of radius $K$, with a total number 
of horizontal bonds equal to $N$ and passing through the origin of the lattice.
The corresponding partition function is then 
\be\label{kih}
{\cal Z}(N,M) \; = \; \sum_{p\in \widetilde{\cal P}_{0}(N,M)} w(p) \; .
\ee
See Figure 4.}
\begin{figure}
\centerline{\epsfbox{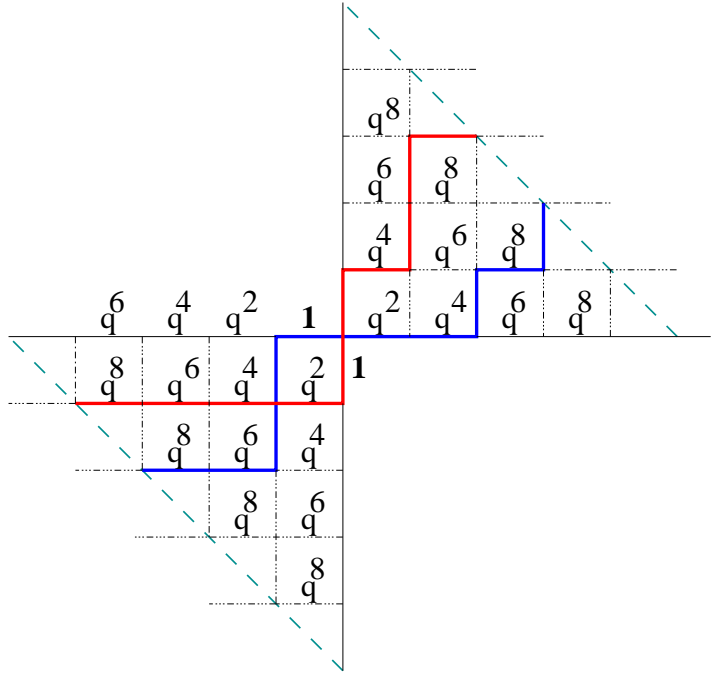}}
\caption{The second path representation.}
\end{figure}
\noindent
{\theorem 
\be
\| \psi_N(-L,K) \|^2 \; = \; Z(N,M) \; = \;  {\cal Z}(N,M)
\ee}
\begin{itemize}
\item[] {\bf Proof.} The first equality comes from Theorem \ref{main}, 
(\ref{bond2}) and (\ref{bond3}). The second
equality is again a consequence of Theorem \ref{main}, (\ref{bond300}) 
and (\ref{bond30}).
\newline
\blackbox
\end{itemize}

\section{Recursion Relations.}

The partition functions ${\cal Z}(N,M)$, defined in (\ref{kih}),
can be related to explicitly known objects such as the
$Z_{q}(n,m)$ given in (\ref{ex}). This can be used effectively in 
numerical or symbolic computations.

{\theorem  The partition function (\ref{kih}) fulfills the following
relation
\be\label{pf}
{\cal Z}(N,M)=\sum_{n+n'=N} Z_{q}(n,K-n) \cdot \left \{ Z_{q}(n'-1,L-n'+1)+
Z_{q}(n',L-n') \right \}\; .
\ee
}
\noindent 
\begin{itemize}
\item[] {\bf Proof.} Consequence of (\ref{genrecq}).
\newline
\blackbox
\end{itemize}
%\begin{figure}\label{pff}
%\centerline{  
%\epsfbox{pf.eps}} 
%\caption{
%The partition function is split in two parts, according
%to whether the spin at position $x=0$ (red bond) is down (figure
%on the left) or up (right). All paths on the upper triangle start at
%the origin and end on the dashed line $n+m=L$. The paths on the lower
%triangle start at $(-1,0)$, when the spins at $x=0$ is down, or
%at $(0,-1)$, when the spins at $x=0$ is up, and end at the
%dashed line $N+M-(n+m)=L+1$ in both cases.
%}
%\end{figure}
%\begin{figure}
%\begin{center}
%\resizebox{!}{7.2 truecm}{\includegraphics{pf.eps}}   
%\vskip .2 cm
%\parbox{10truecm}{\caption{\baselineskip=16 pt\small\label{pff} 
%The partition function is split in two parts, according
%to whether the spin at position $x=0$ (red bond) is down (figure
%on the left) or up (right). All paths on the upper triangle start at
%the origin and end on the dashed line $n+m=L$. The paths on the lower
%triangle start at $(-1,0)$, when the spins at $x=0$ is down, or
%at $(0,-1)$, when the spins at $x=0$ is up, and end at the 
%dashed line $N+M-(n+m)=L+1$ in both cases.}
%}
%\end{center}
%\end{figure}
This expression for the partition function of the model 
can be written in terms of the partition function of a
genuine anisotropic Heisenberg model for $1 \leq x \leq 
N+M$, as we now show. Our first results is:
{\theorem \label{thm:pfav}
The partition function (\ref{pf}) is given by
\be\label{ave}
{\cal Z}(N,M)=Z_{q}(N,M)~\langle q^{-2(K+1)S_{L}} \rangle_{N,M}
\ee
where $Z_{q}(N,M)$ is the partition function of the anisotropic
Heisenberg model, $S_{L}=\sum_{x=-L}^{0} \alpha_{x}$ and the 
symbol $\langle \cdot \rangle$ denotes the expectation value
in the canonical ensemble of the anisotropic Heisenberg model.
}
\begin{itemize}
\item[] {\bf Proof.} We apply 
the property (\ref{TF}) to translate the partition functions in 
(\ref{pf}):
\be\label{trans} 
Z_{q}(n'-1,L+1-n')=q^{-2(K+1)(n'-1)} Z_{q}(N-n'+1,M-L-1+n';N,M),
\ee
In the same way we obtain
\be\label{trans1}
Z_{q}(n',L-n')=q^{-2(K+1)n'} Z_{q}(N-n',M-L+n';N,M)
\ee
for the second term in (\ref{pf}). Substituting (\ref{trans})
and (\ref{trans1}) into (\ref{pf}) we get
\begin{eqnarray}
{\cal Z}(N,M)&=&\sum_{n+n'=N} q^{-2(K+1)n'}Z_{q}(n,K-n) 
\left\{Z_{q}(N-n',M-L+n';N,M)\right. \nonumber\\
&+&\left. q^{2(K+1)} Z_{q}(N-n'+1,M-L-1+n';N,M)\right \}.
\end{eqnarray}
By using the Theorem \ref{TF}, we rewrite the terms between braces 
as follows
\be\label{pf1}
{\cal Z}(N,M)=\sum_{n+n'=N} q^{-2(K+1)n'} Z_{q}(n,K-n)
Z_{q}(n,K-n;N,M).
\ee
The above expression can be interpreted as the average value of 
$q^{-2(K+1)(N-n)}$ over all the paths from the origin to 
$(N,M)$ that pass through the point $(n,m)$. Here, $n$ is the number of
down spins, i.e., horizontal steps in the path, to the left of the pinning site.
As the quantity to be averaged only depends $n$, not on the individual path,
we just need to know the distribution of $n$, which is given by
the probabilities $p_{N,M}(n)$, defined by
\be
p_{N,M}(n)=\frac{Z_q(n,n-N)Z_q(n,n-N;N,M)}
{\sum_{n} Z_q(n,n-N)Z_q(n,n-N;N,M)}
=\frac{Z_q(n,n-N)Z_q(n,n-N;N,M)}{Z_q(N,M)}. 
\ee
%\bea
%p_{(N,M)}(n)&=&\frac{Z_q(n,n-N)Z_q(n,n-N;N,M)}
%{\sum_{n} Z_q(n,n-N)Z_q(n,n-N;N,M)}\nonumber\\
%&=&\frac{Z_q(n,n-N)Z_q(n,n-N;N,M)}{Z_q(N,M)}.
%\eea
Then the result (\ref{ave}) follows and this completes the proof of the 
theorem.
\newline
\blackbox
\end{itemize}
{\theorem
The partition function satisfies
\be\label{rec1}
{\cal Z}(N,M)={\cal Z}(N-1,M)+{\cal Z}(N,M-1)
\ee
}
\begin{itemize}
\item[] {\bf Proof.} Consequence of (\ref{bond2}) and (\ref{genrec}). 
\newline
\blackbox
\end{itemize}
{\theorem
The partition function satisfies
\be\label{rec2}
{\cal Z}(N,M)\; = \; \sum_{n+m=K}Z_{q}(n,m)[Z_q(N-n,M-m-1)+Z_q(N-n-1,M-m)]
\ee
}
\begin{itemize}
\item[] {\bf Proof.} Consequence of (\ref{genrec}) and of the observation
\be
Z(n,m;N,M-1) \; = \; Z_q(N-n,M-m-1)
\ee
and
\be
Z(n,m;N-1,M) \; = \; Z_q(N-n-1,M-m)
\ee
which can be derived from (\ref{bond2}).
\newline
\blackbox
\end{itemize}
\vskip .6 cm
\noindent
\noindent {\large \bf Acknowledgments.}
O.B. thanks FAPESP for support under grant 01/08485-6 and Prof. W. 
Wreszinski. P.C. thanks the Instituto de Fisica da Universidade de Sao 
Paulo, where this work was partially done, and Prof. W. Wreszinski for the 
invitation and warm hospitality. Based on work supported in part by the
National Science Foundation under grant \# DMS-0303316.

\end{document}